*An Astro2020 APC White Paper*
# Investing for Discovery in Astronomy

*Joan Najita, NOAO Chief Scientist, najita@noao.edu*

**Thematic areas:** State of the profession, opportunity and discovery, ground-based OIR

**Summary:**

How should we invest our available resources to best sustain astronomy's thrilling track record of discovery, established over the past few decades? Two strong hints come from (1) our history of astronomical discoveries and (2) literature citation patterns that reveal how discovery and development activities in science are strong functions of team size. These argue that progress in astronomy hinges on support for a diversity of research efforts in terms of team size, research tools and platforms, and investment strategies that encourage risk taking.

These ideas also encourage us to examine the implications of the trend toward "big team science" and "survey science" in astronomy over the past few decades, and to reconsider the common assumption that progress in astronomy always means "trading up" to bigger apertures and facilities. Instead, the considerations above argue that we need a balanced set of investments in small- to large-scale initiatives and team sizes both large and small. Large teams tend to develop existing ideas, whereas small teams are more likely to fuel the future with disruptive discoveries. While large facilities are the "value" investments that are guaranteed to produce discoveries, smaller facilities are the "growth stocks" that are likely to deliver the biggest science bang per buck, sometimes with outsize returns. One way to foster the risk taking that fuels discovery is to increase observing opportunity, i.e., create more observing nights and facilitate the exploration of science-ready data.

## Diverse Paths to Discovery in Astronomy

The exciting discoveries of recent decades illustrates how astronomy remains a youthful field with undiminished potential to surprise us and change the way we think about the Universe. In contrast to fields of research that are dominated by a few important questions or quests, astronomy reaches out in numerous directions simultaneously.

As a result, our discoveries are made in diverse ways. Some result from deliberate searches, while others arise completely unexpectedly. Some may be driven by technological developments, others by human inspiration and/or dogged persistence. Some discoveries are made by large projects and large organized teams, while others are made by small groups or individuals. The observing resources used to make discoveries range from relatively modest to state-of-the-art. Table 1 lists some major astronomical discoveries of the past few decades that have made use of ground-based OIR resources.

**Table 1. Some Major Astronomical Discoveries**

|  | Epoch of discovery | Discovery Resources |
|---|---|---|
| *Discovery of the Kuiper Belt | Jewitt & Luu 1993 | UH 88" |
| *Exoplanets discovered | 1995; Mayor & Queloz (Spitzer 2005 light from exoplanet; 2007 molecules in atm; 2009 weather map) | Mayor & Queloz: 1.9m  Marcy & Butler: 3m Shane and 0.6m CAT. |
| *Black hole at the center of the Milky Way and tests of GR. | 1996-present | 8m VLT, 10m Keck |
| *Dark Energy discovered | 1998 | 2-4m discovery imaging; spectroscopy at Keck, ESO; follow up imaging with HST, CTIO, WIYN, INT, ESO |
| *Dwarf planets in the outer Solar System | 2002 Quaoar; 2005 Eris; 2016 evidence for Planet X | Quaoar Palomar 48"; 4m Blanco/DECam, 8m Subaru/HSC |
| *Measurement of baryon acoustic oscillations, a new cosmological tool | Cole et al. 2005; Eisenstein et al. 2005 | 4m AAO/2dF; 2.5m SDSS |
| Properties and occurrence rates of planets – e.g., super-Earths and Neptunes are common companions | Microlensing, e.g., Cassan et al. 2012 | 1.3m OGLE, 1m PLANET |
| Milky Way companions and merger history of the Milky Way: stellar streams and dwarf galaxies | 1994 Sgr dwarf, etc. | 1.3m 2MASS, 2.5m SDSS, 4m Blanco/DECam |
| Galaxies and quasars beyond z=7, patchiness of reionization | 2011+ | 8-10m spectroscopy, HST, Spitzer; 4m imaging, WISE |
| Optical counterpart to a binary neutron star merger and origin of rare heavy elements | 2017 | Small to large aperture |
| 'Oumuamua: a visiting planetesimal from another Solar System | 2017 | 1.8m PanSTARRS |

* Associated with one or more of these prizes: Bruce Medal, National Medal of Science, Nobel, Shaw Prize in Astronomy, Crafoord Prize, Kavli Prize, Breakthrough Prize.

The table shows that facilities with apertures ranging from small to large have contributed to major discoveries in earlier decades. Major discoveries have been made by teams small and large. The diversity of paths to discovery has been supported by the traditional model of a diverse suite of ground-based OIR telescopes and instruments, both public and private that, taken as a whole, is accessible to a broad community of astronomers, each with their own approach to discovery.

While it is well known that smaller aperture facilities often support research conducted on larger aperture facilities (e.g., imaging on smaller aperture facilities, spectroscopy on larger aperture facilities), the chart also illustrates how relatively small aperture facilities have played the leading role in many major discoveries and the development of new ideas, even when larger aperture facilities are in existence. Perhaps this outcome is a result of the ability to carry out longer-term programs on smaller facilities and to take bigger risks in choosing which scientific challenges to take on.

**What will it take to sustain our track record of discovery in astronomy? Our own historical record argues that it requires support for a diversity of research efforts in terms of team size (large or small), research tools and platforms (e.g., large or small aperture; tried-and-true workhorse instruments or state-of-the art), and support for risk taking.**

The ground-based OIR community is moving ahead with just such a diversity of research approaches and efforts. At the NOAO community meeting "NOAO Community Needs for Science in the 2020s" (https://www.noao.edu/meetings/2020decadal/), designed to assist the US ground-based astronomical community in preparing for the 2020 Decadal Survey, participants described how discovery in the 2020s will be multi-faceted with facilities large and small playing important roles. While we anticipate trail-blazing science from the flagship facilities LSST and JWST, new discoveries will likely be made and horizons opened with optical interferometry (e.g., *CHARA*), high accuracy astrometry (*Gaia*), high cadence photometry (*TESS, CHEOPS, PLATO; ZTF, Blackgem, ATLAS, LCO, ASAS-SN, PanSTARRS*), wide-field imaging (e.g., *DES, Legacy Surveys*) and wide-field spectroscopy (*SDSS-V, DESI, APOGEE, PFS, 4MOST, MOONS*).

A general message from the meeting is that astronomy in the 2020s will be driven by diverse questions, an abundance of new research tools, and openly available data. While future big-ticket items (JWST and its successors, ELTs) will continue to be important for discovery, making advances and discoveries in astronomy will not be contingent on these: a multitude of existing facilities, capably instrumented, will open new horizons, solve problems, and raise new questions. Further details about the meeting are available in a recent white paper (see section 2 of Najita 2019; https://arxiv.org/abs/1901.08605).

### Small Teams Disrupt, Large Teams Develop

A parallel argument about the need to support diverse research approaches comes from a broad literature study reported earlier this year about the effect of team size on the kind of research they produce. In their paper, "Large teams develop and small teams disrupt science and technology," Wu, Wang, & Evans (2019, Nature, 566, 378; https://arxiv.org/pdf/1709.02445.pdf) describe how the citation patterns from 65 million papers, patents, and software, reveal that smaller research teams (3 or fewer) more commonly fuel the future, introducing new ideas and opportunities that disrupt science and technology, whereas large teams more commonly develop existing ideas and opportunities.

Individual authors and small teams tend to dig deeper into the past and build on older, less popular ideas, whereas large teams tend to develop recent, prominent ideas. The effect extends down to the level of individuals, who alter their behavior depending on the size of their group, and are more likely to produce innovative results as they move from large to small teams, down to work as solo investigators. The work of large and small groups is synergistic, with discoveries generated by small research teams often becoming the hot topics that large teams develop.

The authors attribute the difference in behavior and research outcomes to the level of risk tolerance of small and large groups. Solo authors and small teams, with less to lose and more to gain, are more willing to take risks, whereas large groups require a constant funding stream to support their work and are risk-averse as a result. **Because both disruption and development are critical to the advancement of science, Wu et al. argue strongly that science policies should support a diversity of team sizes.**

The way in which discoveries generated by small research teams become topics that large teams develop is reminiscent of the business world, where new ideas and opportunities created by startup companies are purchased for development by large established companies. In business, venture capitalists provide seed funding for startups to ensure a continued stream of new ideas. **In astronomy, access to observing opportunity is the equivalent incubator for small team science.**

This vital resource, small team access to observing opportunities, is potentially under threat from both the continued trend toward big team science in astronomy and the continued quest for ever-larger facilities.

### Big Team Science in Astronomy

"Big team science" is increasingly common in astronomy, with team sizes that are increasingly large. LIGO's detection of gravitational waves took $1B and a team of 1000 people. The team that will carry out the Dark Energy Spectroscopic Instrument (DESI) Survey on the 4-m Mayall telescope at Kitt Peak numbers over 600. More generally, "survey science" is all the rage, with the success of SDSS, PanSTARRS, DES, DECaLS, and many other surveys, both from the ground and in space, over the past few decades.

Large team science (in the form of Key Science Programs) will also be the dominant mode of the US-ELT Program, which seeks to secure federal funding for both GMT and TMT. One concern about ELTs is that despite their high cost (~$1B) they offer the same number of observing nights as smaller telescopes. As a result, only a limited number of ideas can be explored with such facilities. To address this concern, the US-ELT Program anticipates that federal funding will enable large surveys that will tackle major problems of the day. The emphasis on survey science makes economic sense. Not only will the immediate science goals of the surveys engage a large community, but the creation and serving of homogeneous datasets for archival re-use can also generate science opportunities beyond the original proposal goals. That is, many researchers can get something out of the limited observing time that will be available.

Following the logic of Wu et al., the above big team efforts in astronomy come with both advantages and disadvantages. They can command greater resources (observing time and funding) in developing the hot ideas of the present, but their disruption/discovery capability may be hampered by the need to reach consensus within the team and to have their (large) funding and observing proposals approved by their peers. **To balance out this effect, it is critical to also support work by smaller groups.**

### "Diversity Crunch" Underway

Another trend in astronomy over the past few decades is the belief that continued discovery depends on ever larger telescope apertures, the idea that progress means trading up to bigger and bigger facilities. This point of view was articulated in a 2018 *Nature* commentary by Matt Mountain and Adam Cohen. They argued that ever-larger facilities are critical to exploration, and that without increased funding and longer-term planning, the US will cede its leadership in astronomy to Europe. An expansion of funding is required because merely shuttering or divesting from smaller facilities cannot meet even the operational costs of new facilities.

Beyond concerns about how long this perspective can be sustained into the future (Najita 2019; https://arxiv.org/abs/1901.08605), underlying this argument is the belief that smaller facilities have little substantial to offer when larger facilities are available, a view that is contradicted by Table 1. As a result, in previous funding cycles, smaller aperture facilities have been divested in the quest to create funding wedges for larger aperture facilities. These decisions have reduced the capacity for diverse, innovative, risk-tolerant, and small team studies.

As one example, the reduction in NSF support for Kitt Peak telescopes has led to exciting new research missions for these facilities. However, the new missions have a restricted science focus and/or focus on team science. The 4m Mayall is now host to DESI, a forefront "big team science" cosmology project. The federal share of the 3.5m WIYN was used to launch an NSF-NASA research program on exoplanets. The 2.1m has been robotized and currently hosts KPED, the Kitt Peak 84-inch Electron Multiplying CCD Demonstrator, which is optimized to follow up short duration transients and close white dwarf binaries. The latter are potential gravitational wave sources.

As a result, the number of nights available to the broad US community for investigations on any topic is greatly reduced (cf. ACP white paper by S. Ridgway), hampering our disruptive discovery potential.

### Economics of Discovery and Risk Taking

The argument by Cohen & Mountain that we can't close enough small facilities to enable the facilities of the future is a clarion call to reconsider the "bigger is always better" assumption and how we can continue our track record of discovery in astronomy in a cost-constrained environment. Basically, if we can make big discoveries with smaller, existing facilities (as the history of discovery in astronomy shows; Table 1), and these facilities cost much less to operate and upgrade than future facilities, we should continue to invest robustly in the smaller facilities as cost-efficient discovery platforms. These facilities should be also made available to investigators who study diverse topics with diverse team sizes. **That is, it is important to balance investment in ELTs or other future facilities with continued access to smaller facilities by diverse groups.**

Restated in the parlance of the investing world, large facilities are the "value" investments that are guaranteed to produce discoveries, while small facilities are the "growth stocks" that are likely to deliver the biggest science bang per buck, sometimes with outsize returns in truly disruptive discoveries (e.g., discovery of exoplanets, dark matter, outer Solar System). Investments in the latter category are critical because that is where growth arises reliably and at modest cost, an important consideration in a cost-constrained environment.

Another critical element needed for continued discovery in astronomy is support for risk taking in research. Merit-based peer review alone is probably not enough. Wu et al. describe how teams that receive federal funding behave like large groups, tending to be conservative and less discovery-oriented, no matter the team size. They speculate that competition for scarce resources leads to a conservative review process. A similar dynamic is likely at work in the evaluation of observing proposals, especially when the oversubscription rate is high, which limits our ability to support risk taking. The solution is not to discard merit-based peer review, which is a valued egalitarian approach that enables broad participation, but to **encourage risk-taking by reducing proposal pressure, i.e., to create more observing opportunities.**

### NOAO Community Survey

The above ideas find strong synergy in the results of the recent NOAO Community Survey carried out online in May 2019. Advertised to the broad ground-based OIR community, the survey polled the community on their overall priorities for investment in the 2020s in a funding-limited environment. For the nearly 500 survey respondents, their highest priority was for observing time (74% of respondents ranked it "critical"), followed by archival data products (45%), new instrumentation (42%), software and data pipelines (39%), student and mentor training (31%), with new telescopes (27%) rounding out the top six. This ordering makes a lot of sense. The first 5 items are important to do science now. New telescopes take longer to come on line and cost a lot more. These and other survey results will be reported in greater detail in a future issue of the NOAO *Currents* e-newsletter (https://www.noao.edu/currents/).

**Apparently respondents think that new telescopes are important but not the most critical need for the coming decade. Alternatively, if new telescopes are built, their development should not compromise these higher priority needs for scientific success in the 2020s or our ability to make disruptive discoveries.**

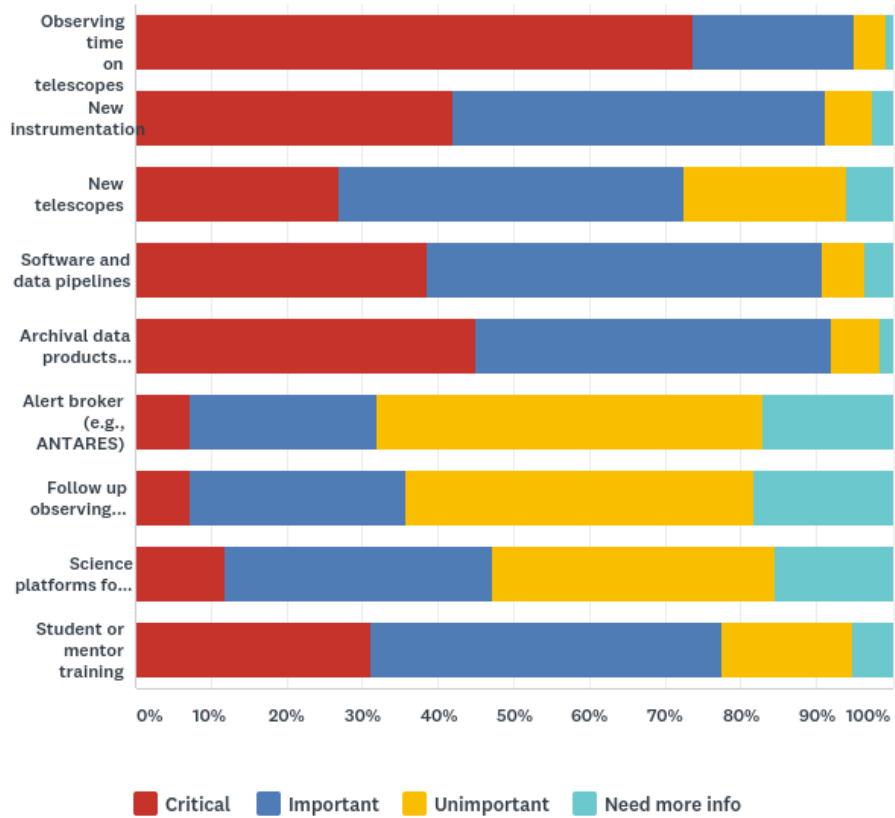

**Figure 1**. Funding priorities from the May 2019 NOAO Community Survey. Respondents were asked to indicate the importance of each resource in a funding-limited environment. The highest priority is (1) **observing time**, followed by (2) archival data products, (3) new instrumentation, (4) software and data pipelines, (5) student or mentor training, and (6) new telescopes.

The survey also found that among instruments and facilities, the demand for ELTs (ranked "very important" by 51% of respondents) is similar to that for highly multiplexed wide-field spectroscopy on ~8-m telescopes (45%) and for the existing 8-m Gemini telescopes. That is, ELTs are important future facilities, but they are one of several options. The diversity of needs and research approaches for ground-based OIR astronomy is an advantage in future planning: they offer options in putting together a balanced investment portfolio. If funding is adequate to support a balanced portfolio that includes ELTs, terrific! They are an extremely valuable component of any portfolio that can afford them, and they will undoubtedly advance astronomy in impressive ways. But if funding is tight, careful thought and agile planning (that is responsive to a changing fiscal situation) is needed to ensure a balanced portfolio.

### Many Revolutions Now…for Less

In assembling a balanced portfolio, modest-aperture facilities offer attractive growth options, especially when recycling is in the picture: many smaller revolutions can happen quickly and nimbly with modest investments, in contrast to the long development times of major facilities. As described in the NOAO Decadal Survey planning meeting and the discoveries of previous decades, new frontiers such as the time domain (via a host of experiments) and the outer Solar System are being explored on the ground with relatively small telescopes.

Equipping older facilities with new instrumentation and repurposing them for new missions (e.g., wide-field imaging such as DECam on 4m Blanco; highly-multiplexed MOS, as in DESI on 4m Mayall, PFS on 8m Subaru and ZTF at Palomar 48"; NEID – high precision RV spectroscopy on 4m WIYN), a trend this past decade, is a strategy that can cut cost and development time. DESI was conceived in early 2009 and will begin commissioning in 2019. The extreme precision RV spectrograph NEID, which builds on state-of-the-art instrumentation experience, has a mere few-year development timeline. With greater multiplexing (wider fields of view, larger format detectors, greater wavelength coverage, more spectra simultaneously), a smaller aperture facility performs like a larger one, at modest cost.

Investments in modest-aperture (2m-8m) facilities can be funded through NSF's Mid-Scale Innovations Program (MSIP). MSIP funding could be used to **buy back open access observing time on NSF facilities previously targeted for divestment** (4m Mayall, 3.5m WIYN, Kitt Peak 2.1m) enabling broad community access to state-of-the-art capabilities. The current uses of these facilities demonstrate that they remain powerful platforms for discovery. MSIP funding could also be directed toward **instrumentation development on non-federal facilities in exchange for open observing nights,** or **straight purchase of observing nights on non-federal facilities.** MSIP funding this past decade has led to open access time with Las Cumbres Observatory, the CHARA array, and with the LLAMAS IFS on Magellan.

Funding for **high-level archival data products and the science platforms that can mine them also creates new science opportunities**, open to a broad cross-section of our community, by allowing us to carry out personalized "observations" using archival data. Investments in support of this possibility offer a relatively low-cost pathway to discovery. MSIP support has previously funded access to data products from ZTF, DES, PFS/Subaru, and Keck All Sky Precision AO.

Looking back at the past few decades, it is clear that the swing toward large team and survey science has led to valuable changes in perspective, e.g., broad recognition of the value of large homogeneous data sets, data pipelines, and software for data mining. Because discoveries made with the resulting archival datasets will typically require follow up observations tailored to specific projects, often in the context of small team science, we now need to rebalance our portfolio to ensure that observing opportunities are broadly available to pursue these discoveries and create new ones.

## Summary and Recommendations

1. Astronomy's continued success requires **rebalanced support for a diversity of research efforts in terms of team size (large or small), research tools and platforms (e.g., large or small aperture; tried-and-true workhorse or state-of-the art instrumentation), and support for risk taking**, as argued by our own historical record of discovery.

2. **Access to observing opportunity is the incubator for small team science in astronomy.** The work of large and small groups is synergistic, with discoveries generated by small research teams often becoming the hot topics that large teams develop. Support for small and large teams is needed to enable both discovery/disruption and development in astronomy, which are critical to the advancement of our field.

3. We need to **balance investment in large facilities (ELTs or other future facilities) and large team science with continued access by smaller groups and individuals to observing opportunities on facilities of all sizes.** Observing opportunities for small teams are potentially under threat from both the continued trend toward big team science in astronomy and the continued quest for ever-larger facilities.

4. **Risk taking, which is critical to continued discovery in astronomy, can be fostered by increasing observing opportunity, i.e., creating more observing nights and facilitating the exploration of science-ready data.** Low cost options including (1) NSF buyback of observing time on federal facilities previously targeted for divestment, (2) use of MSIP to purchase observing time on non-federal facilities, (3) funding for instrument development on non-federal facilities in exchange for open access observing nights, (4) funding for science platforms and efficient access to high-level archival data products that enable unanticipated discoveries.